\documentstyle[12pt]{article}

\makeatletter
\ifcase\@ptsize
  \font\tenmsy=msbm10
  \font\sevenmsy=msbm7
  \font\fivemsy=msbm5
\or
  \font\tenmsy=msbm10 scaled \magstephalf
  \font\sevenmsy=msbm8
  \font\fivemsy=msbm6
\or
  \font\tenmsy=msbm10 scaled \magstep1
  \font\sevenmsy=msbm8
  \font\fivemsy=msbm6
\fi
\newfam\msyfam
\textfont\msyfam=\tenmsy  
\scriptfont\msyfam=\sevenmsy
\scriptscriptfont\msyfam=\fivemsy
\def\Bbb{\ifmmode\let\next\Bbb@\else
\def\next{\errmessage{Use \string\Bbb\space only in math mode}}\fi\next}
\def\Bbb@#1{{\Bbb@@{#1}}}
\def\Bbb@@#1{\fam\msyfam#1}
\newfam\euffam
\font\sixeuf=eufm6
\font\eighteuf=eufm8
\font\twelveeuf=eufm10 scaled\magstep1
\textfont\euffam=\twelveeuf
\scriptfont\euffam=\eighteuf
\scriptscriptfont\euffam=\sixeuf
\def\euf{\fam\euffam\twelveeuf}
\makeatother

\newcommand{\BN}{{\Bbb{N}}}

\newcommand{\BR}{{\Bbb{R}}}
\newcommand{\BZ}{{\Bbb{Z}}}
\newcommand{\Bid}{1\!{\rm l}}

\def\ES{{\euf S}}
\newcommand{\myboldmath}{\boldmath}
\def\pn{\par\noindent}

\def\w{{\cal W}}

\def\be{\begin{equation}}
\def\ee{\end{equation}}
\def\ba{\begin{array}}
\def\ea{\end{array}}
\def\bea{\begin{eqnarray}}
\def\eea{\end{eqnarray}}
\def\bean{\begin{eqnarray*}}
\def\eean{\end{eqnarray*}}
\def\bl{\begin{list}{}{}}

\def\my{$\!$\smiley}

\newcommand{\reseteqn}{\setcounter{equation}{0}}
\newcommand{\mysection}{\reseteqn\section}

\if@twoside
\else
\fi
\def\smiley{
  \begin{picture}(8,10)
    \thinlines        
    \put(5,6){\circle{12}}
    \put(1,1.5){${\scriptscriptstyle\smile}$}
    \put(0,7){.}
    \put(6,7){.}
    \put(5,6){\line(0,-1){2}}
  \end{picture}
}
\begin{document}
  \pagestyle{empty}
  \begin{raggedleft}
IASSNS-HEP-96/54\\
hep-th/9605151\\
May 1996\\
  \end{raggedleft}
%  \begin{center}
%{\eurm DRAFT}
%  \end{center}
  $\phantom{x}$ %%% \vskip 0.618cm\par
  {\LARGE\bf
  \begin{center}
On Fusion Rules\\ 
in\\ 
Logarithmic Conformal Field Theories 
  \end{center}
  }\par
  \vfill
  \begin{center}
$\phantom{X}$\\
{Mic$\hbar$ael A.I.~Flohr\footnote[1]{ email: {\tt flohr@sns.ias.edu}}}\\
$\phantom{X}$\\
{\em School of Natural Sciences\\
Institute for Advanced Study\\
Olden Lane\\
Princeton, NJ 08540, USA}
  \end{center}\par
  \vfill
  \begin{abstract}
  \noindent 
  We find the fusion rules for the $c_{p,1}$ series of logarithmic 
  conformal field theories. This completes our attempts to generalize
  the concept of rationality for conformal field theories to the logarithmic
  case. A novelty is the appearance of negative fusion coefficients which
  can be understood in terms of exceptional quantum group representations.
  The effective fusion rules (i.e.\ without signs and multiplicities) 
  resemble 
  the BPZ fusion rules for the virtual minimal models with conformal grid
  given via $c=c_{3p,3}$. This leads to the conjecture that (almost) all
  minimal models with $c=c_{p,q}$, $(p,q)>1$, belong to the class of
  rational logarithmic conformal field theories.
  \end{abstract}
  \vfill\vfill\vfill
  \newpage
%
%%< INTRODUCTION >%%%%%%%%%%%%%%%%%%%%%%%%%%%%%%%%%%%%%%%%%%%%%%%%%%%%%%%%
%
  \setcounter{page}{1}
  \pagestyle{plain}
  \mysection{Introduction}
  \pn
It is now more or less one do-decade ago since the concept of rationality
of conformal field theory (CFT) made its first appearance through the minimal
models of Belavin, Polyakov and Zamolodchikov \cite{BPZ83}. Since then
rational conformal field theories (RCFTs) became a main tool in modern
theoretical physics.
  \par
More or less one decade later it has been shown \cite{Gur93} that CFTs whose 
correlation functions exhibit logarithmic behavior, can still be consistently 
defined. Recently there has been increasing interest in these logarithmic
conformal field theories (LCFTs). Such LCFTs include the WZNW model on the
supergroup $GL(1,1)$ \cite{RoSa92}, the $c_{p,1}$ models 
\cite{Flo95,GaKa96,Gur93,Kau91,Kau95,Roh96}, gravitationally dressed conformal
field theories \cite{BiKo95,Sch92}, and some critical disordered models 
\cite{CKT95,MaSe96}. They are believed to be important for the description of
certain statistical models, in particular in the theory of (multi-) critical
polymers and percolation \cite{DuSa87,Flo95,Sal92}, in the quantum Hall effect
\cite{Flo96,MaSe96,MiRe96,WeWu94}, and in 2-dimensional turbulence 
\cite{Flo96a,RaRo95}.
They also play a role in the so called unifying $\w$ algebras \cite{BEHHH94}
and they might play a role in the description of normalizable zero modes for 
string backgrounds \cite{EMN96,KoMa95}.
  \par
In our paper \cite{Flo95}, hereafter referred to as \my, we started to
generalize the concept of rationality to the case of LCFTs. There, some 
details have not been resolved in a completely satisfactory way. In 
particular, the Verlinde $S$ matrix seems to have a block structure preventing
one to use the Verlinde formula to get the fusion rules. Moreover, the quantum
group structure, which underlies every RCFT \cite{GoSi91} normally in a
hidden way, becomes visible in the case of LCFTs, where in addition the 
values of the quantum deformation parameters $q_{\pm}$ are of the form
$q_+ = \exp(2\pi i/p)$, $q_- = \exp(2\pi ip) = 1$. The
consequences for the representation theory of LCFTs remained unclear in \my.
  \par
The aim of this paper is to close these remaining gaps and thus complete
our attempts to generalize rationality to the logarithmic case. 
We use the notations of \my, and to save space the interested reader is asked 
to consult our earlier work for reference.
  \par
The flow of this letter is as follows: In section 2 we review the problems
left in \my. Section 3 is devoted to the consequences of exceptional
representations due to the quantum group structure. We explain how the
so called exceptional representations introduce some degree of freedom
into the linear combinations of characters such that both, modular invariant
partition functions and $S$ matrices with ``good'' fusion rules, can be
obtained at the same time. The constraint of integer valued fusion 
coefficients leads to a unique solution for the $S$ matrix. But the linear
combinations of characters do not agree with results from direct examinations
of the representations and their embedding structure 
\cite{GaKa96,GaKa96a}\footnote[1]{We are greatful to M.R.~Gaberdiel and
H.G.~Kausch for pointing this out to us}, but are related to the latter 
by simple linear
combinations with integer coefficients. If instead one uses the 
correct characters, the $S$ matrix seems to yield bad fusion rules. This 
obstacle is resolved by an interesting limiting procedure.
We continue in
section 4 with remarks on the meaning of negative fusion coefficients and
an observation regarding the effective fusion rules (forgetting the signs),
which relates the $c_{p,1}$ LCFT to the would-be minimal model $c_{3p,3}$
(which does not exist since $3p$ and $3$ are not coprime). We conclude with
some speculations on a possible generalization of rational LCFTs for
arbitrary $c_{np,n}$, $n>2$, models.
  \par
We find it remarkably that we obtain simple fusion rules for these ``minimal''
LCFTs, despite the fact that their representation theory has some important
unusual features (e.g.\ indecomposable highest weight \nopagebreak 
representations, non-diagonalizable $L_0$) \cite{GaKa96,Kau95,Roh96}.
  \par
%
%%< SECTION 2 >%%%%%%%%%%%%%%%%%%%%%%%%%%%%%%%%%%%%%%%%%%%%%%%%%%%%%%%%%%%
%
  \mysection{Characters \& Partition Functions}
  \pn
In \my\ modular invariant partition functions have been obtained for the
LCFTs with $c=c_{p,1}$. The surprising fact was that these LCFTs have
a finite closing operator algebra with respect to the maximally extended
chiral symmetry algebra $\w(2,2p-1,2p-1,2p-1)$. Therefore, they also have
a finite set of characters. The building blocks of the characters are
the Dedekind $\eta$ function $q^{1/24}\prod_{n\in\BN}(1-q^n)$ and the
following set of ordinary and affine $\Theta$ functions:
\bea
  \Theta_{\lambda,k}            &=& 
    \sum_{n\in\BZ}q^{(2kn+\lambda^2)/4k}\,,\\
  (\partial\Theta)_{\lambda,k}  &=& 
    \sum_{n\in\BZ}(2kn+\lambda)q^{(2kn+\lambda^2)/4k}\,,\\
  (\nabla\Theta)_{\lambda,k}    &=& i\tau(\partial\Theta)_{\lambda,k} 
    = \frac{1}{2\pi}\log(q)(\partial\Theta)_{\lambda,k}\,.
\eea
Here, $q=\exp(2\pi i\tau)$ is the modular parameter, and the last equation
means that we formally rewrite $2\pi i\tau$ as $\log(q)$, if we deal with
$q$-series expressions.
  \par
A basis for the characters of the $c_{p,1}$ model is given by the following
$3p-1$ linearly independent functions
\bea\label{eq:chars1}
  \chi^{}_{\pm\lambda,p}      &=& \frac{1}{\eta}\Theta_{\lambda,p} \pm
    \frac{1}{\eta} 
    \left[(\partial\Theta)_{\lambda,p}+(\nabla\Theta)_{\lambda,p}\right]\,,\\
  \label{eq:chars2}\tilde{\chi}^{}_{\lambda,p} &=& \frac{\sqrt{2}}{\eta}
    \left[(\partial\Theta)_{\lambda,p}-(\nabla\Theta)_{\lambda,p}\right]\,,
\eea
where $0\leq\lambda\leq p$. Notice, that $\chi_{-p,p}\equiv\chi_{p,p}$, and
$\tilde{\chi}_{0,p}\equiv\tilde{\chi}_{p,p}\equiv 0$. The conformal dimensions
of the corresponding fields are then given by 
\be
  h_{\pm\lambda,p} = \frac{(p\pm(p-\lambda))^2 - (p-1)^2}{4p}\,.
\ee
In \my, all characters $\tilde{\chi}_{\lambda,k}$ got an unphysical
multiplicity of $\sqrt{2}$ in order to obtain a modular invariant partition
function
\be
  Z_{{\rm log}}[p] = |\chi^{}_{0,p}|^2 + |\chi^{}_{p,p}|^2 +
    \sum_{\lambda=1}^{p-1}\left[
    \chi^{}_{\lambda,p}\chi^*_{-\lambda,p} + 
    \chi^{}_{-\lambda,p}\chi^*_{\lambda,p}
    + |\tilde{\chi}^{}_{\lambda,p}|^2\right]\,.
\ee
Moreover, all except 2 of the characters have $\log(q)$ terms. This is
insofar disturbing, as an explicit calculation of the vacuum character
yields $\chi^{}_{{\rm vac},p}=(\Theta_{p-1,p}+(\partial\Theta)_{p-1,p})/
(p\eta)$ without logarithmic terms, in contrast to $\chi^{}_{p-1,p}$. 
Moreover, this basis of characters does not coincide with the physical
one, i.e.\ with characters obtained from direct calculations of
the number of states of the representations. The precise structure of the
representations and the physical characters have been found by M.R.~Gaberdiel 
and H.G.~Kausch \cite{GaKa96,GaKa96a}. 
  \par
In this section, we will use a set of characters suggested by the results
of \my, that promises good modular porperties and, perhaps, to obtain
fusion rules via the Verlinde formula.  
At the end of section 3 we will make contact with the physical 
basis of characters and discuss the differences between the two sets
which turn out to be related by integer valued linear combinations. 
The $S$ matrix for our basis of characters is given by
\be
  S_{(p)} = \left(\ba{c|c}
   \sqrt{\frac{1}{2p}}
   [\cos(\frac{\pi\lambda\lambda'}{p})+\sin(\frac{\pi\lambda\lambda'}{p})]&0\\
   \hline
   0 & -\sqrt{\frac{2}{p}}\sin(\frac{\pi\mu\mu'}{p})
  \ea\right)\,.
\ee
where in the upper left block we have $-p<\lambda,\lambda'\leq p$, and in
the lower right block $0<\mu,\mu'<p$. The block structure of the $S$ matrix
and the multiplicity 2 of the $\tilde{\chi}_{\lambda,p}$ characters are a
hint to the quantum group structure showing up in these LCFTs.
  \par
Actually, we have 2 representations with characters 
$\tilde{\chi}_{\lambda,k}^{\pm}$ such that 
$\tilde{\chi}_{\lambda,k}^{+} + \tilde{\chi}_{\lambda,k}^{-} = \sqrt{2}
\tilde{\chi}^{}_{\lambda,k}$, the latter having vanishing quantum dimension.
This precisely happens in quantum groups, if the quantum deformation parameters
become roots of unity. Then, additional so called exceptional 
representations appear in pairs, whose quantum dimensions add up to 0 
\cite{GoSi91}. The point is that every RCFT has an underlying quantum group 
structure, but precisely for $c=c_{p,1}$, one of the corresponding quantum 
group parameters becomes $\exp(2\pi ip)=1$ and, as already mentioned in \my, 
the quantum group structure becomes visible within the RCFT itself. Notice, 
that the characters $\tilde{\chi}^{}_{\lambda,p}$ have signs in their 
$q$-expansion which is a further hint to an additional quantum number. It is 
now tempting to check whether we can split the representations corresponding 
to the $\tilde{\chi}^{}_{\lambda,k}$ characters in such a way that we obtain 
both, a modular invariant partition function and an $S$ matrix from which we 
can calculate good fusion rules via the Verlinde formula.
  \par
In fact, this is the case. If we split each $\tilde{\chi}^{}_{\lambda,k}$
characters into a pair and redo our analysis of \my\ with a general 
ansatz for all characters 
\bea\label{eq:lincomb}
  \chi^{}_{\lambda,p} &=& \frac{1}{\eta}\left[
  \alpha_{\lambda,p}\Theta_{\lambda,p} + 
  \beta_{\lambda,p}(\partial\Theta)_{\lambda,p} + 
  \gamma_{\lambda,p}(\nabla\Theta)_{\lambda,p}\right]
  \,,\\ \label{eq:lincombtilde}
  \tilde{\chi}^{\pm}_{\mu,p} &=& \frac{1}{\eta}\left[
  \alpha^{\pm}_{\mu,p}\Theta_{\mu,p} +
  \beta^{\pm}_{\mu,p}(\partial\Theta)_{\mu,p} + 
  \gamma^{\pm}_{\mu,p}(\nabla\Theta)_{\mu,p}\right]\,,
\eea
we have more possibilities for writing down a candidate modular invariant 
partition function. One remark is necessary here. We cannot just extend the 
$S$ matrix to the enlarged set of characters, since the enlarged set is no 
longer linear independent. But since the characters are supposed to be split 
in such a way that adding them again yields characters to representations 
with vanishing quantum dimensions, it is sufficient to include only one of 
the two characters, say $\tilde{\chi}^{+}_{\lambda,p}$ into our set for which 
we calculate the $S$ matrix. The $S$ matrix for the set of characters,
where $\tilde{\chi}^{-}_{\lambda,p}$ replaces $\tilde{\chi}^{+}_{\lambda,p}$,
is the same up to some signs (in particular the quantum dimensions 
$S_{{\rm vac}}^{\tilde{\lambda}}/S_{{\rm vac}}^{{\rm vac}}$ change
sign). This will lead to unavoidable signs in the fusion rules which we
will explain in the next section. Another important consequence of this is
that $S$ is no longer unitary in the usual sense. Instead, we have that
$S$ is almost unitary with respect to the metric induced by the 
sesqui-linear form of the partition function,
$Z_{{\rm log}}[p] = \vec{\chi}^t\cdot{\cal N}\cdot\vec{\chi}$,
where ${\cal N}$ is given as
\be
  {\cal N} = \left(\delta_{\lambda,-\lambda'}\right)_{-p<\lambda,\lambda'
  \leq p}\oplus\left(\tilde{\delta}_{\mu,\mu'}\right)_{0<\mu,\mu'<p}\,.
\ee
We then have $S{\cal N}S^{\dagger} = (\Bid_{2p})\oplus(-\tilde{\Bid}_{p-1})$ 
and $S^2 = \Bid_{3p-1}$, $S{\cal N}S^t = {\cal N}$, 
where $\Bid_n$ denotes the $n\times n$ identity matrix, and $\,\tilde{}\,$ 
indicates parts of the matrices which correspond to $\tilde{\chi}$
characters. This reflects the fact that LCFTs do not completely factorize
in left and right chiral part, since otherwise neither conformal invariance
of the 4-point functions nor modular invariance of the partition function
can be assured.
  \par
To understand this behavior, let us introduce the split of characters 
into the partition function. It turns out that one has now two possibilities,
\bea
  Z_{{\rm log}}^+[p] 
         &=& \sum_{\lambda=-p+1}^{p}\chi^{}_{\lambda,p}\chi^*_{-\lambda,p}
             + \sum_{\mu=1}^{p-1}\left[
             |\tilde{\chi}^+_{\mu,p}|^2 + 
             |\tilde{\chi}^-_{\mu,p}|^2\right]\,,\\
  Z_{{\rm log}}^-[p] 
         &=& \sum_{\lambda=-p+1}^{p}|\chi^{}_{\lambda,p}|^2
             + \sum_{\mu=1}^{p-1}\left[
             \tilde{\chi}^+_{\mu,p}\tilde{\chi}^{-^{\scriptstyle *}}_{\mu,p}+
             \tilde{\chi}^-_{\mu,p}\tilde{\chi}^{+^{\scriptstyle *}}_{\mu,p}
             \right]\,.
\eea
Both cases are modular invariant for appropriate chosen linear combinations
(\ref{eq:lincomb}), (\ref{eq:lincombtilde})
of the characters. Extending the $S$ matrix by hand (the prescription for
this can be found in \cite{MMS88}) to incorporate the
split characters, we find the surprising result
$S_{{\rm ext}}{\cal N}^+S^{\dagger}_{{\rm ext}} = {\cal N}^-$ and vice versa, 
where still $S_{{\rm ext}}^2=\Bid_{4p-2}$. Thus, the action of $S_{{\rm ext}}$
intertwines $Z_{{\rm log}}^+[p]$ and $Z_{{\rm log}}^-[p]$. 
A closer look shows that $Z_{{\rm log}}^+[p]$ and $Z_{{\rm log}}^-[p]$ 
just differ by the sign of the $\log(q\bar q)$ part in their
$q$-series expansion. Since $\log(q\bar q) = 2\pi i(\tau - \bar{\tau})$, we
see that successive application of $S:\tau\mapsto\frac{-1}{\tau}$ and
$S^{\dagger}:\tau\mapsto\frac{-1}{\bar{\tau}}$ precisely yields this sign.
  \par
{}From the above we conclude that it is not possible to find a good extended
$S$ matrix. In fact, one can show that the method of \cite{MMS88} does not
work in the case, where representations are split into {\em different\/}
characters, instead of having just a multiplicity $>1$ in the partition
function.
We will see later that it is possible to find a unitary (not extended) $S$ 
matrix, if we incorporate the non trivial sesqui-linear form ${\cal N}$ 
in the proper way. The
problem is that our characters yield an $S$ matrix with $S^2=\Bid$, which does
not depend on the choice of a basis. What we would like is a modified matrix
$\ES$ such that $\ES\ES^{\dagger}=\Bid$, $\ES^2={\cal N}$, i.e.\
${\cal N}$ gives rise to non trivial charge conjugation.
  \par
So far, the correct coefficients of the linear combinations 
(\ref{eq:lincomb}), (\ref{eq:lincombtilde})
are not yet uniquely determined. Every quadruple $(\chi^{}_{\lambda,p},
\chi^{}_{-\lambda,p},\tilde{\chi}^+_{\lambda,p},\tilde{\chi}^-_{\lambda,p})$
is determined up to three of its coefficients (namely $\gamma_{\lambda,p}=
-\gamma_{-\lambda,p}, \gamma^{\pm}_{\lambda,p}$), if we impose the condition 
of minimal integer coefficients for the $q,\bar q$-series expansion of the 
partition function. All other constants, i.e.\ all constants of terms
without $\log(q)$, assume the values as given by (\ref{eq:chars1}),
(\ref{eq:chars2}). The remaining free constants are determined by 
requiring integer valued fusion rules.
  \par
%
%%< SECTION 3 >%%%%%%%%%%%%%%%%%%%%%%%%%%%%%%%%%%%%%%%%%%%%%%%%%%%%%%%%%%%
%
  \mysection{{\myboldmath $S$} Matrix \& Fusion Rules}
  \pn
We can now either determine the remaining coefficients by directly 
considering the fusion rules \underline{{\bf (case I)}}, by just requiring 
that the $S$ matrix should be symmetric \underline{{\bf (case II)}}, or by 
matching the characters with explicit calculations of the spectra of the 
representations \cite{GaKa96,GaKa96a,Kau95,Roh96}
\underline{{\bf (case III)}}. It was pointed out by M.R.~Gaberdiel and
H.G.~Kausch, see \cite{GaKa96a}, that the characters of the solutions to
case I and case II disagree with the direct calculations. 
In ordinary RCFTs one would
expect that all these approaches yield the same solution. This seems to
be no longer true in the logarithmic case. But nonetheless, the different
solutions are consistent with each other.
  \pn
\underline{{\bf Getting ``good'' fusion rules (case I):}}
In case I, we obtain one pair of 
solutions which is valid for all $p>1$, namely $\gamma_{\lambda,p} = 0$, 
$\gamma^+_{\lambda,p} = 1$ or $\gamma_{\lambda,p} = 2$, 
$\gamma^+_{\lambda,p} = 3$ for all $0 < \lambda < p$, where the solution 
with $\gamma_{\lambda,p} = 0$ is more appealing, as we have mentioned
in the introduction: all characters $\chi^{}_{\lambda,p}$ are then free from
$\log(q)$ terms. In case II, we also obtain a solution, namely
$\gamma_{\lambda,p} = 0$, $\gamma^+_{\lambda,p} = \pm i\sqrt{3}$, for 
all $0 < \lambda < p$. Unfortunately, this solution does not work if $p$ 
is divisible by 3 due to zeros in the row/column of the $S$ matrix 
corresponding to the vacuum character, such that the Verlinde formula 
becomes singular. Nevertheless, the second solution has some significance, 
as will be pointed out in the last section. Case I and case II can be 
viewed as specializations of the more general requirement that the $S$ 
matrix fulfills $S_j^k = \pm S^{\dagger\,j}_{\,k}$. If one makes the 
(reasonable) assumption that $\gamma_{\lambda,p} = \pm\gamma_{\mu,p}$, 
$\gamma^+_{\lambda,p} = \pm\gamma^+_{\mu,p}$ for all $0 < \lambda,\mu < p$ 
one obtains a general expression 
$\gamma_{\lambda,p}^+ = \pm f_{\pm}(\gamma_{\lambda,p})$, where $f_{\pm}$ 
are rational functions involving fractional powers. The sign label 
corresponds to the choice of (anti-) symmetry for the $S$ matrix entries.
  \par
As an example we consider the $c=c_{2,1} = -2$ model. With the 
requirements from the last section we obtain the $S$ matrix
\be
  S_{(2)} = \left(\begin {array}{ccccc} 
        \frac{i(xy-1)}{2(x+y)} & \frac{1}{2} & \frac{i(1-xy)}{2(x+y)} &
          -\frac{1}{2} & -\frac{i(x^2+1)}{x+y} \\
        \frac{1}{2} & \frac{1}{2} & \frac{1}{2} & \frac{1}{2} & 0 \\ 
        \frac{i(1-xy)}{2(x+y)} & \frac{1}{2} & \frac{i(xy-1)}{2(x+y)} &
          -\frac{1}{2} & \frac{i(x^2+1)}{x+y} \\ 
        -\frac{1}{2} & \frac{1}{2} & -\frac{1}{2} & \frac{1}{2} & 0 \\
        \frac{i(y^2+1)}{x+y} & 0 & -\frac{i(y^2+1)}{x+y} & 0 &
          \frac{i(1-xy)}{x+y}
  \end {array}\right)\,,
\ee
where we abbreviated $x = \gamma_{1,2}$ and $y = -\gamma^+_{1,2}$. 
The general solutions
$y=f(x)$ are $y=\pm\sqrt{-2x^2-3}$ and $y=\pm\sqrt{2x^2+1}$. One can easily 
check that the solutions for $x,y$ mentioned above yield integer valued fusion
rules via the Verlinde formula
\be\label{eq:verlinde}
   N_{ij}^k=\sum_r\frac{S_i^r S_j^r (S^{-1})_{r}^{k}}{S_{{\rm vac},r}}\,,
\ee
where $S_{{\rm vac},r}$ denotes the row corresponding to the vacuum 
representation. Notice, that the fusion coefficients may be negative.
Let us denote the $\w$ conformal families corresponding to the characters
$\chi^{}_{\lambda,p}$, $\tilde{\chi}^{}_{\lambda,p}$, and
$\tilde{\chi}^{\pm}_{\lambda,p}$ by $[h_{\lambda,p}]$, 
$[\tilde h_{\lambda,p}]$, and $[\tilde{h}^{\pm}_{\lambda,p}]$ respectively.
Choosing $y=-1,x=0$, we get for our example the following fusion rules 
\be\label{eq:fusel1}
  \begin{array}{rclcl}
    {}           [0]&\times&[0]           &=&[0]                         \,,\\
    {}           [0]&\times&[-\frac{1}{8}]&=&[-\frac{1}{8}]              \,,\\
    {}           [0]&\times&[\frac{3}{8}] &=&[\frac{3}{8}]               \,,\\
    {}           [0]&\times&[1]           &=&[1]                         \,,\\
    {}           [0]&\times&[\tilde{0}]   &=&[\tilde{0}]                 \,,\\
    {}[-\frac{1}{8}]&\times&[-\frac{1}{8}]&=&[0]-[\tilde{0}]             \,,\\
    {}[-\frac{1}{8}]&\times&[\frac{3}{8}] &=&[1]+[\tilde{0}]             \,,\\
    {}[-\frac{1}{8}]&\times&[1]           &=&[\frac{3}{8}]               \,,
  \end{array}
  \begin{array}{rclcl}
    {}[-\frac{1}{8}]&\times&[\tilde{0}]   &=&[-\frac{1}{8}]-[\frac{3}{8}]\,,\\
    {} [\frac{3}{8}]&\times&[\frac{3}{8}] &=&[0]-[\tilde{0}]             \,,\\
    {} [\frac{3}{8}]&\times&[1]           &=&[-\frac{1}{8}]              \,,\\
    {} [\frac{3}{8}]&\times&[\tilde{0}]   &=&[\frac{3}{8}]-[-\frac{1}{8}]\,,\\
    {}           [1]&\times&[1]           &=&[0]                         \,,\\
    {}           [1]&\times&[\tilde{0}]   &=&-[\tilde{0}]                \,,\\
    {}   [\tilde{0}]&\times&[\tilde{0}]   &=&[1]-[0]+3[\tilde{0}]        \,.
  \end{array}
\ee
The choices $y=1,x=0$, or $y=\pm 2,x=\pm 3$, $y=\pm 2, x=\mp 3$ yield the
same fusion rules up to some possible signs which correspond to exchanging
$\tilde{\chi}^+$ with $\tilde{\chi}^-$. From this we can now read off 
the fusion rules where the $[\tilde{h}_{\lambda,p}]$ representations 
(in our example there is just $[\tilde{h}_{1,2}] = [\tilde{0}]$) are split. 
Giving only the non trivial cases we have
\be\label{eq:fusel2}
  \begin{array}{rclcl}
    {}              [0]&\times&[\tilde{0}^{\pm}]&=&[\tilde{0}^{\pm}]     \,,\\
    {}   [-\frac{1}{8}]&\times&[-\frac{1}{8}]   &=&[0]+[\tilde{0}^-]     \,,\\
    {}   [-\frac{1}{8}]&\times&[\frac{3}{8}]    &=&[1]+[\tilde{0}^+]     \,,\\
    {}   [-\frac{1}{8}]&\times&[\tilde{0}^+]    &=&[-\frac{1}{8}]        \,,\\
    {}   [-\frac{1}{8}]&\times&[\tilde{0}^-]    &=&[\frac{3}{8}]         \,,\\
    {}    [\frac{3}{8}]&\times&[\frac{3}{8}]    &=&[0]+[\tilde{0}^-]     \,,
  \end{array}
  \begin{array}{rclcl}
    {}    [\frac{3}{8}]&\times&[\tilde{0}^+]    &=&[\frac{3}{8}]         \,,\\
    {}    [\frac{3}{8}]&\times&[\tilde{0}^-]    &=&[-\frac{1}{8}]        \,,\\
    {}              [1]&\times&[\tilde{0}^{\pm}]&=&[\tilde{0}^{\mp}]     \,,\\
    {}[\tilde{0}^{\pm}]&\times&[\tilde{0}^{\pm}]&=&[0]+2[1] 
                                                   +2[\tilde{0}^+]       \,,\\
    {}[\tilde{0}^{\pm}]&\times&[\tilde{0}^{\mp}]&=&2[0]+[1]
                                                   +2[\tilde{0}^-]       \,.
  \end{array}
\ee
For the last two equations one has to take into account that $[1]-[0]=
([0]+2[1])-(2[0]+[1])$ and $-[\tilde 0^{\mp}] = [\tilde 0^{\pm}]$.
We note that the solution $y=\pm i\sqrt{3},x=0$ yields a unitary $S$ matrix
and (up to signs) again the same fusion rules, but without multiplicities,
i.e.\ all $N_{ij}^k \in \{-1,0,1\}$. For completeness we also give the
$T$ matrix which is {\em non-diagonal}, a general feature of LCFTs. Again
putting $x=0,y=-1$ we obtain
\be T_{(2)} = \left(
  \ba{ccccc}
    e^{\pi i/6} &               &             &                & \\
                & e^{-\pi i/12} &             &                & \\
                &               & e^{\pi i/6} &                & \\
                &               &             & e^{11\pi i/12} & \\
    -\frac{1}{2}e^{\pi i/6} &   & \frac{1}{2}e^{\pi i/6} &     & e^{\pi i/6}
  \ea\right)\,,
\ee
which fulfills $(ST)^3 = \Bid$. The generalization of the $T$ matrix for
arbitrary $p$ is obvious, the rows for the $[\tilde{h}_{\lambda,p}]$ 
representations get off-diagonal entries with $\pm\frac{1}{2}$ the value of
the diagonal in the columns to the $[h_{\pm\lambda,p}]$ representations, i.e.
\be
  T_{(p)} = \left(e^{2\pi i(h_{\lambda,p}-c/24)}\delta_{\lambda,\lambda'}
          + e^{2\pi i(h_{\mu-p,p}-c/24)}(\delta_{\mu,\mu'}
            + \frac{1}{2}\delta_{\mu-p,\lambda'} 
            - \frac{1}{2}\delta_{\mu-p,-\lambda'}) 
            \right)_{-p<\lambda,\lambda'\leq p,\,p<\mu,\mu'<2p}\,.
\ee
Here, the $\mu,\mu'$ labels indicate the rows and columns to the 
$[\tilde{h}_{\mu,p}]$ representations, the $\lambda,\lambda'$ labels refer to
the $[h_{\lambda,p}]$ representations.
  \par
In general, case I yields (up to some irrelevant signs) the following
expression for the $S$ matrix:
\be\label{eq:smatp} S_{(p)} =
  \left(\ba{cccc|c}
    \sqrt{\frac{1}{2p}}\exp(\frac{\pi i\lambda\lambda'}{p}) & & & & 
        i\sqrt{\frac{2}{p}}\sin(\frac{\pi\mu\mu'}{p}) \\ \cline{5-5}
    &\phantom{00}& & & 0 \ldots 0 \\ \cline{5-5} 
    & & \sqrt{\frac{1}{2p}}\exp(\frac{\pi i\lambda\lambda'}{p}) & &
        -i\sqrt{\frac{2}{p}}\sin(\frac{\pi\mu\mu'}{p}) \\ \cline{5-5}
    & & & & 0 \ldots 0 \\ \hline
    \hfill\vline & 0 \hfill\vline & \hfill\vline & 0 & \\
    -i\sqrt{\frac{2}{p}}\sin(\frac{\pi\mu\mu'}{p})\phantom{x}\hfill\vline & 
        \vdots \hfill\vline &
        i\sqrt{\frac{2}{p}}\sin(\frac{\pi\mu\mu'}{p}) \hfill\vline & 
        \vdots &
        -i\sqrt{\frac{2}{p}}\sin(\frac{\pi\mu\mu'}{p}) \\
    \hfill\vline & 0 \hfill\vline & \hfill\vline & 0 &
  \ea\right)\,.
\ee
Here, $-p <\lambda,\lambda' \leq p$ and in all blocks $0 < \mu,\mu' < p$.
Therefore, the $S$ matrix has an interesting structure. It consists of
blocks which by itself are well known. Indeed, the blocks 
\be
  i\sqrt{\frac{2}{p}}\left(\sin(\frac{\pi\mu\mu'}{p})\right)_{0<\mu,\mu'<p}
\ee
are nothing else than (up to an irrelevant overall factor $-i$) the $S$ matrix 
of the current algebra $A^{(1)}_1$ at level $p+2$. The other block,
\be
  \sqrt{\frac{1}{2p}}\left(\exp(\frac{\pi i\lambda\lambda'}{p})\right)_{-p<
    \lambda,\lambda'\leq p}
\ee
resembles (up to the same factor $-i$) the $S$ matrix of a $c=1$ model with
compactification radius $2R^2 = p$. Consequently, these blocks yield good
fusion rules by themselves denoted by ${\cal S}_{\lambda\mu}^{\nu}$ and
${\cal E}_{\lambda\mu}^{\nu}$ respectively. Thus, it is easy to see that
$S_{(p)}$ yields integer valued fusion rules which actually turn out to be
linear combinations of the form $N_{ij}^{k} = a{\cal E}_{ij}^{k} + 
b{\cal S}_{ij}^{k}$ with $a,b\in\{-1,0,1\}$. 
  \pn
\underline{{\bf Getting a ``good'' {\boldmath $S$} matrix (case II):}}
Let us next discuss the solution of case II. As a matter of fact,
putting $\gamma^+_{\lambda,p} = \pm i\sqrt{3}$, $\gamma_{\lambda,p}=0$ 
always yields real, symmetric, unitary $S$ matrices, which we denote by 
$S^{\pm}_{(p)}$. Unfortunately, they have vanishing
entries in some rows and columns, in particular the row and column to the
vacuum representation, if and only if $p$ is divisible by $3$. One can show
that this problem can partially be overcome, if one considers the matrix
$\ES_{(p)} = \frac{1}{\sqrt{2}}(\sqrt{i}S^+_{(p)}+\sqrt{-i}S^-_{(p)})$,
which is the average of the two $S$ matrices to the split characters
$\tilde{\chi}^+_{\lambda,p}$ and $\tilde{\chi}^-_{\lambda,p}$ respectively.
The precise form of this average is dictated by the conditions 
$\ES\ES^{\dagger}=\Bid$ and $\ES^2 = {\cal N}'$ with
\be
  {\cal N}' = \left(\delta_{\lambda,-\lambda'}\right)_{-p<\lambda,\lambda'
  \leq p}\oplus\left(-\tilde{\delta}_{\mu,\mu'}\right)_{0<\mu,\mu'<p}\,.
\ee
The charge conjugation might seem unusual, but since 
$\tilde{\chi}^+_{\lambda,p} = -\tilde{\chi}^-_{\lambda,p}$, we get exactly 
${\cal N}'$ if $C(\tilde{\chi}^+_{\lambda,p}) = \tilde{\chi}^-_{\lambda,p}$
and vice versa. 
  \par
If one calculates fusion rules with the help of the $\ES$ matrix, one
finds for $p\not=3p'$ integer valued fusion coefficients $N_{ij}^k\in
\{-1,0,1\}$. If $p=3p'$, one finds instead rational fusion coefficients
$N_{ij}^k\in\{0,\pm\frac{1}{3},\pm\frac{2}{3},\pm 1\}$, which is a bit
disturbing. The origin of these factors of $\frac{1}{3}$ might be understood
as follows:
  \par
It has been pointed out in \my\ that the field content of the $c_{p,1}$
LCFTs can be read off from the conformal grid of $c_{3p,3}$. The mapping is
$h_{\lambda,p}=h(3p,3)_{1,p-\lambda}$, $h_{-\lambda,p}=h(3p,3)_{1,3p-\lambda}$
and $\tilde{h}_{\lambda,p}=h(3p,3)_{1,p+\lambda}$. Naively, this
``minimal model'' does not exist \cite{BPZ83}, because $3p$ and $3$ are
certainly not coprime. On the other hand, the famous BPZ fusion rules for 
a minimal model with $c=c_{p,q}=1-6\frac{(p-q)^2}{pq}$ and fields 
$\phi_{r,r'}$ of conformal dimensions $h(p,q)_{r,r'} = \frac{1}{4pq}\left[
(pr'-qr)^2 - (p-q)^2\right]$ are
\be\label{eq:bpz}
  {}[h(p,q)_{r,r'}]\times[h(p,q)_{s,s'}] = 
  \sum_{{n=|r-s|+1\atop n+r+s-1\equiv 0\,{\rm mod}\,2}}^{{\rm min}(q-1,r+s-1)}
  \sum_{{n'=|r'-s'|+1\atop n'+r'+s'-1\equiv 0\,{\rm mod}\,2}}^{
    {\rm min}(p-1,r'+s'-1)}[h(p,q)_{n,n'}]
\ee
and are well defined for arbitrary $p,q$. In the case that $p,q$ are coprime,
the $S$ matrix for the characters of the minimal model with $c=c_{p,q}$
is \cite{CIZ87}
\be
  S(p,q)_{(r,r')}^{(s,s')} = (-)^{rs'+r's+1}2\sqrt{\frac{2}{pq}}
    \sin\left(\pi\frac{p}{q}rs\right)\sin\left(\pi\frac{q}{p}r's'\right)\,,
\ee
where $1\leq r,s\leq q-1$ and $1\leq r',s'\leq p-1$. Let us now assume that
$p=\alpha^{{\rm\#}p}p'$, $q=\alpha^{{\rm\#}q}q'$ with $p',q'$ coprime. 
Of course, one of the powers is 1, but it is convenient to use this symmetric 
notation. Then we can define a matrix
\be\label{eq:smat}
  S(p,q)_{(r,r')}^{(s,s')} = (-)^{rs'+r's+1}2\sqrt{\frac{2}{pq}}
    \sin\left(\pi\frac{p}{\alpha^{{\rm\#}p}q}rs\right)
    \sin\left(\pi\frac{q}{\alpha^{{\rm\#}q}p}r's'\right)\,.
\ee
One can now prove that this $S$ matrix yields BPZ fusion rules 
(\ref{eq:bpz}) for arbitrary $p,q$. The surprise is that $S(3p,3)$ induces 
the
fusion rules of our LCFTs with $c=c_{p,1}$, if we forget about all signs
and multiplicities. We call such fusion rules {\em effective\/} fusion rules.
Moreover, ${\ES_{(p)}}_{k}^{l}$ equals $S(3p,3)_k^l$ up to phases (correct
relabeling implicitly understood), if
$p$ is not divisible by 3. In this case ${\rm\#}(3p)={\rm\#}3=1$, $\alpha=3$. 
The above formula (\ref{eq:smat}) shows that for $p$ divisible by 3, 
extra powers have to be introduced into the denominator of one of the sine 
terms. 
  \pn
\underline{{\bf Getting ``good'' characters (case III):}}
The solution given above does yield good fusion rules but has the 
disadvantage that the basis of characters is still not the one observed
in physics, i.e.\ which one gets by explicitly considering representations
and counting states. The latter task has been dealt with in 
\cite{GaKa96,Kau95,Roh96}. The result is
\bea
  \chi^{}_{0,p}      & = & \frac{1}{\eta}\Theta_{0,p}\,,\\
  \chi^{}_{p,p}      & = & \frac{1}{\eta}\Theta_{p,p}\,,\\
  \chi^+_{\lambda,p} & = & \frac{1}{p\eta}\left[(p-\lambda)\Theta_{\lambda,p}
                       + (\partial\Theta)_{\lambda,p}\right]\,,\\
  \chi^-_{\lambda,p} & = & \frac{1}{p\eta}\left[\lambda\Theta_{\lambda,p}
                       - (\partial\Theta)_{\lambda,p}\right]\,,\\
  \tilde{\chi}^+_{\lambda,p} & = & \frac{1}{\eta}\left[\Theta_{\lambda,p}
                       + i\alpha\lambda(\nabla\Theta)_{\lambda,p}\right]\,,\\ 
  \tilde{\chi}^-_{\lambda,p} & = & \frac{1}{\eta}\left[\Theta_{\lambda,p}
                       - i\alpha(p-\lambda)(\nabla\Theta)_{\lambda,p}\right]
                       \,,
\eea 
where $0<\lambda<p$. The conformal weights are (in the same order) 
$h(p,1)_{1,p}$, $h(p,1)_{1,2p}$, $h(p,1)_{p-\lambda}$, $h(p,1)_{3p-\lambda}$,
and $h(p,1)_{p+\lambda}$. One easily sees that the partition function
\bea
  \lefteqn{Z_{{\rm log}}[p,\alpha] 
    = |\chi^{}_{0,p}|^2 + |\chi^{}_{p,p}|^2
    + \sum_{\lambda=1}^p\left[
    \chi^+_{\lambda,p}{\tilde{\chi}^{+^{\scriptstyle *}}_{\lambda,p}} +
    {\chi^{+^{\scriptstyle *}}_{\lambda,p}}\tilde{\chi}^+_{\lambda,p} +
    \chi^-_{\lambda,p}{\tilde{\chi}^{-^{\scriptstyle *}}_{\lambda,p}} +
    {\chi^{-^{\scriptstyle *}}_{\lambda,p}}\tilde{\chi}^-_{\lambda,p}\right]
  }\\
  & = &\frac{1}{\eta\eta^*}\left\{|\Theta_{0,p}|^2 + |\Theta_{p,p}|^2 +
    \sum_{\lambda=1}^{p} \left[ 2|\Theta_{\lambda,p}|^2 
    +i\alpha\left((\partial\Theta)_{\lambda,p}(\nabla\Theta)_{\lambda,p}^*
    -(\partial\Theta)_{\lambda,p}^*(\nabla\Theta)_{\lambda,p}\right)
    \right]\right\}\nonumber
\eea
is modular invariant for all $\alpha\in\BR$. Notice the important fact that
the partition function remains modular invariant even for $\alpha=0$ and 
then equals the standard $c=1$ Gaussian model partition function 
$Z(\sqrt{p/2})$. This in particular means that we have a modular invariant
partition function even in the case where the characters do not form a
closed finite dimensional representation of the modular group by themselves.
One can now again calculate an $S$ matrix for one set of linear independent
characters, e.g.\ $\{\chi^{}_{0,p},\chi^{}_{p,p},\chi^{\pm}_{\lambda,p},
[(p+x-\lambda)\tilde{\chi}^+_{\lambda,p}+
(\lambda-x)\tilde{\chi}^-_{\lambda,p}]\}$. 
This $S$ matrix and the fusion rules obtained via the Verlinde
formula (\ref{eq:verlinde}) depend on the value of $\alpha$. Clearly,
the $S$ matrix becomes singular for $\alpha\rightarrow 0$, but the fusion
rules remain well defined. The limes $\alpha\rightarrow 0$ just puts several
of the fusion coefficients to zero. It turns out that in general only the 
fusion rules in the limit $\alpha\rightarrow 0$ are consistent and integer 
valued.
  \par
As an example let us again consider the case $c=c_{2,1}=-2$. The $S$ matrix
reads
\be
  S_{(2,\alpha)} = \left(\ba{ccccc}
    \frac{1}{2\alpha} & \frac{1}{4} & \frac{1}{2\alpha} & -\frac{1}{4} &
      -\frac{1}{4\alpha} \\
    1 & \frac{1}{2} & 1 & \frac{1}{2} & 0 \\
    -\frac{1}{2\alpha} & \frac{1}{4} & -\frac{1}{2\alpha} & -\frac{1}{4} &
      \frac{1}{4\alpha} \\
    -1 & \frac{1}{2} & -1 & \frac{1}{2} & 0 \\
    -2\alpha & 1 & 2\alpha & -1 & 0
  \ea\right)\,.
\ee
The $S$ matrix $S_{(p,\alpha)}$ in general is neither symmetric nor unitary
but again fulfills $S_{(p,\alpha)}^2 = \Bid$. The general expression for the
$S$ matrix is cumbersome, but can easily obtained by applying an appropriate 
base change of characters to, e.g., the matrix $S_{(p)}$ given in 
(\ref{eq:smatp}). Defining $N^i_{jk}(\alpha)$ via the Verlinde  
formula (\ref{eq:verlinde}), the fusion rules are obtained by the limes
$N^i_{jk} = \lim_{\alpha\rightarrow 0}N^i_{jk}(\alpha)$. Care must then 
be taken in interpreting the right hand sides of fusion products: Since
the characters are non longer linearly independent in this limit, we
may have character identities which translate to relations among the
representations. For $\alpha=0$ one has at least that
$2[h(p,1)_{1,p-\lambda}]+2[h(p,1)_{1,3p-\lambda}] = [h(p,1)_{1,p+\lambda}]$,
$0<\lambda<p$, e.g.\ the relation $[\tilde 0] = 2[0] + 2[1]$ in
our example. The interpretation of signs remains the same as before.
The fusion rules now read
\be
  \begin{array}{rclcl}
    {}           [0]&\times&[\Phi]        &=&[\Phi]                      \,,\\
    {}[-\frac{1}{8}]&\times&[-\frac{1}{8}]&=&2[0]+2[1] = [\tilde{0}]     \,,\\
    {}[-\frac{1}{8}]&\times&[\frac{3}{8}] &=&2[0]+2[1] = [\tilde{0}]     \,,\\
    {}[-\frac{1}{8}]&\times&[1]           &=&[\frac{3}{8}]               \,,\\
    {}[-\frac{1}{8}]&\times&[\tilde{0}]   &=&2[-\frac{1}{8}]+2[\frac{3}{8}]
                                             \,,\\
    {} [\frac{3}{8}]&\times&[\frac{3}{8}] &=&2[0]+2[1] = [\tilde{0}]     \,,
  \end{array}
  \begin{array}{rclcl}
    {} [\frac{3}{8}]&\times&[1]           &=&[-\frac{1}{8}]              \,,\\
    {} [\frac{3}{8}]&\times&[\tilde{0}]   &=&2[-\frac{1}{8}]+2[\frac{3}{8}]
                                             \,,\\
    {}           [1]&\times&[1]           &=&[0]                         \,,\\
    {}           [1]&\times&[\tilde{0}]   &=&4[0]+4[1]-[\tilde{0}] 
                                             = [\tilde{0}]               \,,\\
    {}   [\tilde{0}]&\times&[\tilde{0}]   &=&8[0]+8[1] = 4[\tilde{0}]    \,.
  \end{array}
\ee
Again, we can reintroduce the split representations $[\tilde{h}^{\pm}]$ back
into the fusion rules, though their characters coincide in the limit
$\alpha\rightarrow 0$. This yields the following modifications
\be\label{eq:gaka}
  \begin{array}{rclcl}
    {}[-\frac{1}{8}]&\times&[-\frac{1}{8}]      &=&[\tilde{0}^+]     \,,\\
    {}[-\frac{1}{8}]&\times&[\frac{3}{8}]       &=&[\tilde{0}^-]     \,,\\
    {} [\frac{3}{8}]&\times&[\frac{3}{8}]       &=&[\tilde{0}^+]     \,,
  \end{array}
  \begin{array}{rclcl}
    {}           [1]&\times&[\tilde{0}^{\pm}]   &=&[\tilde{0}^{\mp}] \,,\\ 
    {}[\tilde{0}^{\pm}]&\times&[\tilde{0}^{\pm}]&=&2[\tilde{0}^+]
                                                   +2[\tilde{0}^-]   \,,\\
    {}[\tilde{0}^{\pm}]&\times&[\tilde{0}^{\mp}]&=&2[\tilde{0}^+]
                                                   +2[\tilde{0}^-]   \,.  
  \end{array}
\ee
It is a simple task to check that these fusion rules are ``compatible'' with
the ones given before, (\ref{eq:fusel1}), (\ref{eq:fusel2}), if the correct
identifications are taken into account. For instance, we have 
$[0]_I+[\tilde 0^-]_I = [\tilde 0^+]_{III}$, 
$[1]_I+[\tilde 0^+]_I = [\tilde 0^-]_{III}$ in our example, relating the
base within the representations of case I to the base in case III.
  \pn
\underline{{\bf Combining the results:}}
To summarize, we found several possibilities to get the fusion rules
of LCFTs from the modular behavior of the partition function and its
building blocks, the characters. Since there is no unique basis of
characters in the case of LCFTs (due to the integer valued differences
between conformal weights of several representations), several possible
choices of a basis arose. What we called cases I and II above used bases in 
which the fusion rules can be obtained from the full characters, including
logarithmic terms, such that these characters form a closed finite dimensional
representation of the modular group. But these bases are not the physical 
ones. In case III we used the physical basis of characters, but could obtain
fusion rules via Verlinde formula only in the limit of vanishing logarithmic
terms. Although the partition function remains modular invariant in this limit,
the characters do not longer form a finite dimensional representation of the
modular group. Fortunately, the different fusion rules turned out to
be compatible with each other. Although, case III deals with the physical
characters, its fusion rules are difficult to read due to relations among
the representations in the limit of vanishing logarithmic terms. The best
way to obtain the fusion rules is therefore to combine case I and case III.
The recipe is to calculate the fusion rules via case I (where the logarithmic
terms prohibit ``obscuring'' relations) and then make a
transformation into the physical base of case III. This still might yield
ambiguities, since the appearance of negative fusion coefficients allows
the freedom of $n[h] = (n+k)[h] - k[h]$. These must be removed by hand through
consistency requirements. We checked explicitly
all models up to $p=6$ and obtained compatible and consistent results.
In particular, these results agree with direct calculations in \cite{GaKa96a}.
  \par
%
%%< SECTION 4 >%%%%%%%%%%%%%%%%%%%%%%%%%%%%%%%%%%%%%%%%%%%%%%%%%%%%%%%%%%%
%
  \mysection{On The Space of {\boldmath $2$}-dimensional Field Theories}
  \pn  
We do not
understand up to now, whether our view of the $c_{p,1}$ LCFTs as certain 
``minimal models for non coprime $3p,3$'' has any deeper meaning. But we 
emphasize the striking fact that all the structures known for RCFTs and in
particular for the minimal models have counterparts in rational LCFTs.
By finding characters, partition functions, and -- finally -- the fusion
rules for LCFTs, we complete our attempts to generalize {\em rationality\/} 
to LCFTs. Since (\ref{eq:smat}) is valid for arbitrary $p,q$, we conjecture 
that there should exist a whole class of generalized (logarithmic) CFTs, for
which our $c_{p,1}$ LCFTs are just the first examples. For instance, one 
might think of enlarged minimal models (thereby no longer minimal in the 
precise meaning of this term) ${\cal M}(\alpha p,\alpha q)$ with $p,q$ 
coprime and central charge still $c_{p,q}$. The $c_{p,1}$ LCFTs are then 
just the enlarged models ${\cal M}(3p,3)$, where simply the ``minimal'' 
models with ${\cal M}(p,1)$ are empty.
  \par
We remark that the enlarged models ${\cal M}(2p,2q)$ can be interpreted
as $N=1$ supersymmetric extensions of the $N=0$ minimal models 
${\cal M}(p,q)$, where the free Grassman parts have been canceled (otherwise,
the central charge had to be shifted by $+\frac{1}{2}$).
We remind the reader that for unitary minimal $N=1$ supersymmetric models
$p-q = 2$, and that the Ramond sector contains only for $p,q$ even a globally
supersymmetric invariant ``vacuum'' state. Is it just a coincidence that
the model ${\cal M}(3\cdot 2,3\cdot 1)$ with $c=-2$ has a hidden $N=2$
supersymmetry \cite{Sal92}? It might be worthwhile to investigate further 
how supersymmetric and logarithmic CFTs are related. This might shed some
light on the representation theory of $N=2$ supersymmetric conformal field
theories (SCFTs) in general and more specifically the question of the 
relationship between rationality and unitarity for $N=2$ SCFTs. We conjecture 
that non-unitary $N=2$ SCFTs can only be rational, if they are 
{\em logarithmic\/} CFTs, since all attempts so far to find  decompositions 
of their partition functions into finitely many products of characters, i.e.\
$Z=\sum_{i,j\in\Lambda} \chi^*_i{\cal N}_{i,j}\chi^{}_j$ with 
$|\Lambda|<\infty$ and $\chi_i$ modular forms of weight 0, have failed
\cite{EhGa96}. 
The $c=-2$ model would be the first example of such a case, namely the
non-free part of an $N=2$ SCFT with central charge (shifted by the free
Grassmann contribution by $+2$) $c=0$. 
  \par
The characters of the LCFTs studied so far are mixed expressions of
modular forms of weight 0 and 1. We call such theories LCFTs of degree 1.
There are also indications that one can even further conjecture that
in general non-unitary rational $N=2k$ SCFTs are logarithmic of degree $k$.
This might explain the appearance of certain modular weights in 
partition functions of such models, which have been obtained without 
considering the theories as LCFTs. Work in this direction will be reported
elsewhere \cite{Flo00}
  \par
There is an alternative view of the $c_{p,1}$ LCFTs, which might illuminate
our understanding of the general space of 2-dimensional field theories 
(2dFTs), and how CFTs are embedded in this larger space. At least the series
of $c_{p,1}$ LCFTs can be understood as limiting points of certain series
of non-unitary minimal models.
  \par
Let us consider the two sequences ${\cal M}((\alpha+1)p,\alpha q)$ and
${\cal M}(\alpha p,(\alpha+1)q$, with $\alpha\in\BZ_+$. If $\alpha\rightarrow
\infty$, we see that the central charges tend to
\be
  \lim_{\alpha\rightarrow\infty}
    1-6\frac{((\alpha+1)p-\alpha q)^2}{\alpha(\alpha+1)pq} = 
  \lim_{\alpha\rightarrow\infty}
    1-6\frac{(\alpha p-(\alpha+1)q)^2}{\alpha(\alpha+1)pq} =
  1 - 6\frac{(p-q)^2}{pq} = c_{p,1}\,,
\ee
while the {\em effective\/} central charges, defined as $c_{{\rm eff}} = 
c - 24h_{{\rm min}}$, $h_{{\rm min}}$ being the smallest highest weight,
tend to
\be
  \lim_{\alpha\rightarrow\infty} 1 - 6\frac{1}{\alpha(\alpha+1)pq} = 1\,,
\ee
in agreement with the fact that the $c_{p,1}$ LCFTs have effective
central charge $c_{{\rm eff}} = 1$ (see \my). Let us consider the
characters of these minimal models,
\bea
  \chi^{((\alpha+1)p,\alpha q)}_{r,s} & = & \frac{1}{\eta}\left[
    \Theta_{(\alpha+1)pr - \alpha qs,\alpha(\alpha+1)pq} -
    \Theta_{(\alpha+1)pr + \alpha qs,\alpha(\alpha+1)pq}\right]\,,\\
  \chi^{(\alpha p,(\alpha+1)q)}_{r,s} & = & \frac{1}{\eta}\left[
    \Theta_{\alpha pr - (\alpha+1)qs,\alpha(\alpha+1)pq} -
    \Theta_{\alpha pr + (\alpha+1)qs,\alpha(\alpha+1)pq}\right]\,.
\eea
If $\alpha\rightarrow\infty$, the level of the $\Theta$ functions approaches 
arbitrarily close a number containing a square factor, $\alpha^2 pq$. In
such cases, we have the following identity
\be\label{eq:square}
  \sum_{\mu=0}^{\alpha-1}\Theta_{\alpha(\lambda + 2k\mu),\alpha^2 k}
    = \Theta_{\lambda,k}\,,
\ee
which actually is valid for arbitrary values of $k,\lambda$, but makes sense
as an identity between modular forms only for $k,\lambda\in\BZ/2$. This
identity allows us in the limit to (approximately) collect characters  
of the huge minimal models (i.e.\ having a huge field content $\Lambda$ with 
its number $|\Lambda|$ proportional to $\alpha^2$) to a very much smaller 
set. But we must be very careful with our limiting procedure. Considering 
the difference between the two series, we see that
\be
  \Theta_{(\alpha+1)pr - \alpha qs,\alpha(\alpha+1)pq} -
    \Theta_{\alpha pr - (\alpha+1)qs,\alpha(\alpha+1)pq} 
    \stackrel{\alpha\gg 1}{\longrightarrow}
  \frac{pr-qs}{\alpha}\frac{\partial}{\partial(\alpha(pr-qs))}
    \Theta_{\alpha(pr-qs),\alpha(\alpha+1)pq}\,,
\ee
which certainly vanishes in the limit. But we have to renormalize this 
expression to assure integer coefficients in the power series, otherwise
we cannot treat it as a correction term for {\em finite\/} $\alpha$. The
renormalization factor is usually chosen to be $(\partial\Theta)_{\lambda,k} 
= \frac{2k}{2\pi i\tau}\partial_{\lambda}\Theta_{\lambda,k}$, such that we
recover the affine $\Theta$ functions. Plugging this into (\ref{eq:square}),
we obtain (by choosing an appropriate sequence of pairs $r,s$ such that
$\lambda + 2pq\mu = pr-qs$)
\be
 \frac{2\alpha^2 pq}{2\pi i\tau}\sum_{\mu=0}^{\alpha-1}
   \frac{1}{\alpha}\frac{\partial}{\partial(\alpha\lambda)}
   \Theta_{\alpha(\lambda + 2pq\mu),\alpha^2 pq} =
 \frac{2pq}{2\pi i\tau}\frac{\partial}{\partial\lambda}\sum_{\mu=0}^{\alpha-1}
   \Theta_{\alpha(\lambda + 2pq\mu),\alpha^2 pq}
   = (\partial\Theta)_{\lambda,pq}\,. 
\ee
Therefore, we conclude that the characters for the theory at the limit point
have additional terms proportional to $(\partial\Theta)_{\lambda,pq}$. The
different non-trivial factorizations $k=pq$ yield the different non-diagonal
partition functions $Z_{{\rm log}}[p/q]$ for $c_{pq,1}$ models, given in \my.
This is not to be confused with the case where our two series of minimal
models approach ${\cal M}(p,q)$. In this latter case we expect that the
limit point actually is the LCFT ${\cal M}(3p,3q)$. 
  \par
This draws a new picture for the space of general 2dFTs. We certainly have
${\rm 2dFTs}\supset{\rm CFTs}\supset{\rm RCFTs}$. The border between the space
of CFTs and the space of non-conformal 2dFTs appears to be made out of 
LCFTs, which have as a subspace the border between RCFTs and non-conformal
2dFTs, namely the hereby established class of rational LCFTs. In a formal way
we might write ${\rm LCFTs} = \partial\overline{{\rm CFTs}}$. Since the
Zamolodchikov metric is known to remain regular at this border, LCFTs might
serve as a new tool for describing transitions between different CFTs and
also the more general transitions from a CFT into the non-conformal region
and vice versa.
  \bigskip\pn
{\bf Acknowledgment:} I would like to thank  
W.~Eholzer, C.~Nayak, F.~Rohsiepe and in particular V.~Gurarie 
for numerous illuminating discussions and comments. I am especially grateful
to M.R.~Gaberdiel and H.G.~Kausch for sharing with me their insights into 
the representation theory of LCFTs.
This work has been supported by the Deutsche Forschungsgemeinschaft.
% \newpage      
%
%%< REFERENCES >%%%%%%%%%%%%%%%%%%%%%%%%%%%%%%%%%%%%%%%%%%%%%%%%%%%%%%%%%%
%
  

\begin{thebibliography}{l99}
  {\footnotesize\setlength{\itemsep}{0cm}
   \bibitem{BPZ83} {\sc A.A. Belavin, A.M. Polyakov, A.B. Zamolodchikov},
%     {\em Infinite conformal symmetry in two-dimensional quantum field 
%     theory},
      Nucl. Phys. {\bf B241} (1984) 333
   \bibitem{BiKo95} {\sc A. Bilal, I.I. Kogan},
%     {\em On gravitational dressing of 2D field theories in chiral gauge},
      Nucl. Phys. {\bf B449} (1995) 569, {\sf hep-th/9503209}
%!!\bibitem{Blu93} {\sc R. Blumenhagen},
%!!%  {\em $N=2$ Supersymmetric $\w$-Algebras},
%!!   Nucl. Phys. {\bf B405} (1993) 744,\\
%!!   {\sc R. Blumenhagen, R. H{\"u}bel},
%!!%  {\em A Note on Representations of $N=2$ ${\cal SW}$-Algebras},
%!!%  Universit"at Bonn Preprint BONN-TH-94-08 (1994), 
%!!   Mod. Phys. Lett. {\bf A9} (1994) 3193, {\sf hep-th/9407068}
%!!\bibitem{BEHH92} {\sc R. Blumenhagen, W. Eholzer, A. Honecker,
%!!   R. H{\"u}bel},
%!!%  {\em New $N=1$ Extended Superconformal Algebras with Two and Three 
%!!%  Generators},
%!!   Int. Jour. Mod. Phys. {\bf A7} (1992) 7841
   \bibitem{BEHHH94} {\sc R. Blumenhagen, W. Eholzer, A. Honecker, 
      K. Hornfeck, R. H{\"u}bel},
%     {\em Unifying $\w$-Algebras},
      Phys. Lett. {\bf B332} (1994) 51-60, {\sf hep-th/9404113},
%     {\em Coset Realization of Unifying $\w$-Algebras},
%     INFN Turin Preprint DFTT-25-94 (1994),  
      Int. Jour. Mod. Phys. {\bf A10} (1995) 2367, {\sf hep-th/9406203}
   \bibitem{BFKNRV91} {\sc R. Blumenhagen, M. Flohr, A. Kliem, W. Nahm, 
      A. Recknagel, R. Varnhagen},
%     {\em $\w$-Algebras with two and three Generators},
      Nucl. Phys. {\bf B361} (1991) 255
%!!\bibitem{Car86} {\sc J.L. Cardy}, 
%!!%  {\em Operator Content of Two-Dimensional Conformally Invariant Theories}
%!!   Nucl. Phys. {\bf B270} (1986) 186
   \bibitem{CIZ87} {\sc A. Cappelli, C. Itzykson, J.-B. Zuber},
%     {\em Modular invariant partition functions in two dimensions}, 
      Nucl. Phys. {\bf B280}[FS18] (1987) 445, 
%!!%  {\em The A-D-E classification of minimal and $A_{1}^{(1)}$ conformal 
%!!%  invariant theories}, 
      Commun. Math. Phys. {\bf 113} (1987) 1
   \bibitem{CKT95} {\sc J.-S. Caux, I.I. Kogan, A.M. Tsvelik},
      {\em Logarithmic Operators and Hidden Continuous Symmetry in Critical
      Disordered Models},
      preprint OUTP-95-62 (1995), {\sf hep-th/9511134}
%!!\bibitem{DVV88} {\sc R. Dijkgraaf, E. Verlinde, H. Verlinde},
%!!%  {\em $c=1$ Conformal Field Theories on Riemann Surfaces},
%!!   Commun. Math. Phys. {\bf 115} (1988) 649
%!!\bibitem{DoFa84} {\sc V.S. Dotsenko, V.A. Fateev},
%!!%  {\em Conformal Algebra and Multipoint Correlation Functions in 2d 
%!!%  Statistical Models}, 
%!!   Nucl. Phys. {\bf B249}[FS12] (1984) 312,
%!!%  {\em Four-Point Correlation Functions and the Operator Algebra in 2d 
%!!%  Conformal Invariant Theories with Central Charge $c\leq 1$}, 
%!!   Nucl. Phys. {\bf B251}[FS13] (1985) 691,
%!!%  {\em Operator Algebra of Two-Dimensional Conformal Theories with
%!!%  Central Charge $c \leq 1$},
%!!   Phys. Lett. {\bf B154} (1985) 291
   \bibitem{DuSa87} {\sc B. Duplantier, H. Saleur},
%     {\em Exact critical properties of two-dimensional dense selfavoiding 
%     walks},
      Nucl. Phys. {\bf B290}[FS20] (1987) 291
   \bibitem{EFHHV93} {\sc W. Eholzer, M. Flohr, A. Honecker, 
      R. H{\"u}bel, R. Varnhagen},
      {\em $\w$-Algebras in Conformal Field Theory}, in
      {\em Superstrings and Related Topics}, Proc. Trieste Workshop (1993), 
      E. Gava, A. Masiero, K.S. Narain, S. Randjbar-Daemi, Q. Shafi (eds.),
      World Scientific, p.435
   \bibitem{EhGa96} {\sc W. Ehozler, M.R. Gaberdiel},
      {\em Unitarity of rational $N=2$ superconformal theories},
      preprint DAMTP-96-06 (1996), {\sf hep-th/9601163}
   \bibitem{EHH93b} {\sc W. Eholzer, A. Honecker, R. H{\"u}bel},
%     {\em How complete is the classification of $\w$-symmetries?},
      Phys. Lett. {\bf B308} (1993) 42, {\sf hep-th/9302124}
%!!\bibitem{Eho00} {\sc W. Eholzer, N.-P. Skoruppa},
%!!   {\em Modular Invariance and Uniqueness of Conformal Characters},
%!!%  Universit{\"a}t Bonn Preprint BONN-TH-94-16, Max-Planck-Institut f{\"u}r
%!!%  Mathematik Bonn Preprint MPI-94-67, hep-th/9407074 (1994)
%!!   preprint BONN-TH-94-16, MPI/94-67 (1994), to appear
%!!   in Commun. Math. Phys., {\sf hep-th/9407074} 
   \bibitem{EMN96} {\sc J. Ellis, N.E. Mavromatos, D.V. Nanopoulos},
      {\em $D$ Branes from Liouville Strings},
      preprint ACT-04/96, CERN-TH/96-81, CTP-TAMU-11/96, OUTP-96-15P (1996),
      {\sf hep-th/9605046}
%!!\bibitem{FeFu82} {\sc B.L. Feigin, D.B. Fuks},
%!!%  {\em Invariant skew-symmetric differential operators on the line and 
%!!%  Verma modules over the Virasoro algebra},
%!!   Funkt. Anal. Appl. {\bf 16} (1982) 114
%!!\bibitem{FeFu83} {\sc B.L. Feigin, D.B. Fuks},
%!!%  {\em Verma Modules over the Virasoro Algebra}, 
%!!   Funct. Anal. Appl. {\bf 17} (1983) 241,
%!!%  {\em Verma Modules over the Virasoro Algebra}, 
%!!   in {\em Topology}, Proc. Leningrad 1982, L.D. Faddeev, A.A. Mal'cev 
%!!   (eds.),
%!!   Lect. Notes Math. {\bf 1060} (1984) 230, Springer Verlag
%!!\bibitem{Fel89} {\sc G. Felder},
%!!%  {\em BRST Approach to Minimal Models}, 
%!!   Nucl. Phys. {\bf B317} (1989) 215,
%!!   {\em Erratum}, 
%!!   Nucl. Phys. {\bf B324} (1989) 548
%!!\bibitem{FFK89} {\sc G. Felder, J. Fr{\"o}hlich, G. Keller},
%!!%  {\em Braid Matrices and Structure Constants for Minimal Conformal 
%!!%  Models}, 
%!!   Commun. Math. Phys. {\bf 124} (1989) 647
   \bibitem{Flo93} {\sc M. Flohr},
%     {\em $\w$-Algebras, New Rational Models and the Completeness of the
%     $c=1$ Classification},
      Commun. Math. Phys. {\bf 157} (1993) 179, {\sf hep-th/9207019}
   \bibitem{Flo94} {\sc M. Flohr},
%     {\em Curiosities at Effective $c = 1$},
      Mod. Phys. Lett. {\bf A9} (1994) 1071, {\sf hep-th/9312097}
   \bibitem{Flo95} {\sc M. Flohr},
      {\em On Modular Invariant Partition Functions of Conformal Field 
      Theories with Logarithmic Operators},
      preprint CSIC-IMAFF-42-1995 (1995), 
      Int. J. Mod. Phys. {\bf A11} (1996) in press, {\sf hep-th/9509166}
   \bibitem{Flo96} {\sc M. Flohr},
      Mod. Phys. Lett. {\bf A11} (1996) 55-68, {\sf hep-th/9605152}
   \bibitem{Flo96a} {\sc M. Flohr},
      {\em 2-dimensional turbulence: yet another conformal field theory 
      solution},
      to appear
%!!\bibitem{PhD} {\sc M. Flohr},
%!!   {\em The Rational Conformal Quantum Field Theories in Two Dimensions 
%!!   with Effective Central Charge $c_{{\rm eff}} \leq 1$},
%!!   preprint BONN-IR-94-11, (Ph.D. thesis in german, 1994)
   \bibitem{Flo00} {\sc M. Flohr},
      work in preparation
%!!\bibitem{FHW93} {\sc M.D. Freeman, K. Hornfeck, P. West},
%!!%  {\em Commuting quantities and exceptional $\w$-algebras},
%!!   Int. J. Mod. Phys. {\bf A8} (1993) 909
%!!\bibitem{FrWe93} {\sc M.D. Freeman, P. West},
%!!   {\em On the relation between integrability and infinite-dimensional
%!!   algebras}, 
%!!   preprint KCL-TH-93-1 (1993), {\sf hep-th/9303119}
   \bibitem{GaKa96} {\sc M.R. Gaberdiel, H.G. Kausch},
      {\em Indecomposable Fusion Products}, 
      preprint DAMTP 96-36 (1996), {\sf hep-th/9604026}
   \bibitem{GaKa96a} {\sc M.R. Gaberdiel, H.G. Kausch}, 
      {\em A Rational Logarithmic Conformal Field Theory},
      preprint DAMTP-96-54 (1996), to appear
   \bibitem{GoSi91} {\sc C. Gomez, G. Sierra},
%     {\em A Note on Liouville Theory and the Uniformization of Riemann 
%     Surfaces}, in
%     {\em Quantum Field Theory, Statistical Mechanics, Quantum Groups and 
%     Topology}, Proc. Miami (1991), pp.115-122,
      Phys. Lett. {\bf B225} (1991) 51, Int. J. Mod. Phys. {\bf A6} (1991) 
      2045-2074
   \bibitem{Gur93} {\sc V. Gurarie},
%     {\em Logarithmic Operators in Conformal Field Theory},
      Nucl. Phys. {\bf B410} (1993) 535, {\sf hep-th/9303160}
%!!\bibitem{KaPe84} {\sc V.G. Kac, D.H. Peterson},
%!!%  {\em Infinite-Dimensional Lie Algebras, Theta Functions and 
%!!%  Modular Forms}, 
%!!   Adv. Math. {\bf 53} (1984) 125
   \bibitem{Kau91} {\sc H.G. Kausch},
%     {\em Extended Conformal Algebras Generated by a Multiplet of Primary 
%     Fields},
      Phys. Lett. {\bf B259} (1991) 448
   \bibitem{Kau95} {\sc H.G. Kausch},
      {\em Curiosities at $c=-2$}, 
      preprint DAMTP 95-52 (1995), {\sf hep-th/9510149}
%!!\bibitem{KaWa91} {\sc H.G. Kausch, G.M.T. Watts},
%!!%  {\em A Study of $\w$-Algebras using Jacobi Identities},
%!!   Nucl. Phys. {\bf B354} (1991) 740
   \bibitem{KoMa95} {\sc I.I. Kogan, N.E. Mavromatos},
      {\em World-Sheet Logarithmic Operators and Target Space Symmetries
      in String Theory},
      preprint OUTP-95-50 P (1995), {\sf hep-th/9512210}
   \bibitem{MaSe96} {\sc Z. Maassarani, D. Serban},
      {\em Non-Unitary Conformal Field Theory and Logarithmic Operators
      for Disordered Systems},
      preprint SPHT-T96/037 (1996), {\sf hep-th/9605062}
   \bibitem{MMS88} {\sc S.D. Mathur, S. Mukhi, A. Sen},
%     {\em On the Classification of Rational Conformal Field Theories}
      Phys. Lett. {\bf B213} (1988) 303,
%  \bibitem{MMS89} {\sc S.D. Mathur, S. Mukhi, A. Sen},
%     {\em Reconstruction of Conformal Field Theories from Modular Geometry
%     on the Torus},
      Nucl. Phys. {\bf B318} (1989) 483
   \bibitem{MiRe96} {\sc M. Milovanovi\'c, N. Read},
      {\em Edge excitations of paired fractional quantum Hall states},
      preprint Yale Univ. (1996), {\sf cond-mat/9602113}
%!!\bibitem{Nah91} {\sc W. Nahm},
%!!%  {\em A proof of modular invariance},
%!!   Int. J. Mod. Phys. {\bf A6} (1991) 2837, in Proc. Trieste July 1990 
%!!   {\em Topological methods in quantum field theories}, 
%!!   World Scientific, 1991
   \bibitem{RaRo95} {\sc M.R. Rahimi Tabar, S. Rouhani},
%     {\em Turbulent Two Dimensional Magnetohydrodynamics and Conformal
%     Field Theory},
      Ann. Phys. {\bf 246} (1996) 446-458, {\sf hep-th/9503005},
      {\em The Alf'ven Effect and Conformal Field Theory},
      preprint Sharif Univ. Tehran (1995), {\sf hep-th/9507166}
   \bibitem{Roh96} {\sc F. Rohsiepe},
      {\em Non-unitary Representations of the Virasoro Algebra with 
      non-trivial Jordan blocks},
      preprint BONN-IR-xx, (diploma thesis in german, 1996)
   \bibitem{RoSa92} {\sc L. Rozansky, H. Saleur},
      Nucl. Phys. {\bf B389} (1993) 365-423, {\sf hep-th/9203069}
   \bibitem{Sal92} {\sc H. Saleur},
%     {\em Polymers and percolation in two dimensions and twisted $N=2$ 
%     supersymmetry},
      Nucl. Phys. {\bf B382} (1992) 486-531, {\sf hep-th/9111007 }
%     {\em Geometrical lattice models for $N=2$ supersymmetric theories in
%     two dimensions},
      Nucl. Phys. {\bf B382} (1992) 532-560, {\sf hep-th/9111008}
   \bibitem{Sch92} {\sc Ch. Schmidhuber},
      Nucl. Phys. {\bf B404} (1993) 342, {\sf hep-th/9212075};
      Nucl. Phys. {\bf B453} (1995) 156, {\sf hep-th/9506118}
   \bibitem{WeWu94} {\sc Xiao-Gang Wen, Yong-Shi Wu, 
      Yasuhiro Hatsugai},
%     {\em Chiral Operator Product Algebra and Edge Excitations of a 
%     Fractional Quantum Hall Droplet},
      Nucl. Phys. {\bf B422}[FS] (1994) 476-494, {\sf cond-mat/9311038,
      cond-mat/9310027}%,\\
%!!%  {\sc Xiao-Gang Wen, Yong-Shi Wu},\\
%!!%  {\em Chiral Operator Product Algebra Hidden in Certain Fractional 
%!!%  Quantum Hall Wave Functions},
%!!%  Massachusetts Institute of Technology, Physics Department, 
%!!%  Preprint 1994, {\sf cond-mat/9310027}
     } %%% end footnotesize %%%
  \end{thebibliography}
  \end{document}